\documentclass[conference]{IEEEtran}
\IEEEoverridecommandlockouts
\usepackage{cite}
\usepackage{amsmath,amssymb,amsfonts}
\usepackage{algorithmic}
\usepackage{graphicx}
\usepackage{textcomp}
\usepackage{xcolor}
\usepackage{url}
\usepackage{threeparttable}
\usepackage{mathptmx}

\def\BibTeX{{\rm B\kern-.05em{\sc i\kern-.025em b}\kern-.08em
    T\kern-.1667em\lower.7ex\hbox{E}\kern-.125emX}}
\begin{document}

\title{PEAK SHIFT ESTIMATION\\\emph{ - Novel method to estimate ranking of examination data with selective lack - }\\ A Preprint\\
    \thanks{We thank Michelle Pascoe, PhD, from Edanz Group (https://en-author-services.edanz.com/ac) for editing a draft of this manuscript.}
}

\author{\IEEEauthorblockN{Satoshi Takahashi}
    \IEEEauthorblockA{
        \textit{Kanto Gakuin University}\\
        Kanagawa, Japan \\
        0000-0002-1067-6704}
    \and
    \IEEEauthorblockN{Masaki Kitazawa}
    \IEEEauthorblockA{
        \textit{Kitazawa Tech, Rikkyo University}\\
        Kanagawa, Japan \\
        0000-0002-6352-0164}
    \and
    \IEEEauthorblockN{Ryoma Aoki}
    \IEEEauthorblockA{
        \textit{Tokyo Institute of Technology}\\
        Kanagawa, Japan }
    \and
    \IEEEauthorblockN{Atsushi Yoshikawa}
    \IEEEauthorblockA{
        \textit{Tokyo Institute of Technology}\\
        Kanagawa, Japan \\
        0000-0001-7020-5085}
}

\maketitle

\begin{abstract}
    In this paper, we focus on examination results when examinees selectively skip examinations, to compare the difficulty levels of these examinations. We call the resultant data ``selectively omitted examination data.'' An example of this type of examination is university entrance examinations. In their selection of and application to universities, high school students consider the difficulty levels of entrance examinations and their academic abilities. Students skip university entrance examinations that are too difficult or too easy relative to their academic abilities. Moreover, high schools generally release the number of accepted students for each examination but do not publicize the number of students who were not accepted. We can learn the number of students accepted for each examination and organization but not the examinees' identity. Selectively omitted examinations such as certification examinations and the outcome of students' job-hunting activities are archived, and universities usually track students' job-hunting activities and the students' subsequent first professional position. However, no research has focused on this type of data. When we know the difficulty level of these examinations, we can obtain a new index to assess organization ability, how many students pass, and the difficulty of the examinations. This index would reflect the outcomes of their education corresponding to perspectives on examinations. Many methods such as Classical Test Theory and Item Response Theory have been proposed to compare the difficulty levels of examinations or items in the examinations. However, selectively omitted examination data does not meet the requirements of these methods, and cannot be applied to the data. Therefore, we propose a novel method, Peak Shift Estimation, to estimate the difficulty level of an examination based on selectively omitted examination data. First, we apply Peak Shift Estimation to the simulation data and demonstrate that Peak Shift Estimation estimates the rank order of the difficulty level of university entrance examinations very robustly. Peak Shift Estimation is also suitable for estimating a multi-level scale for universities, that is, A, B, C, and D rank university entrance examinations. We apply Peak Shift Estimation to real data of the Tokyo metropolitan area and demonstrate that the rank correlation coefficient between difficulty level ranking and true ranking is 0.844 and that the difference between 80 percent of universities is within 25 ranks. Additionally, we find that one university would entirely reduce the accuracy of the estimation regardless of its true rank and that universities create several communities with the same external factors, such as medical departments and girls' schools, which would decrease the overall accuracy of the estimation. The accuracy of Peak Shift Estimation is thus low and must be improved; however, this is the first study to focus on ranking selectively omitted examination data, and therefore, one of our contributions is to shed light on this method.
\end{abstract}

\begin{IEEEkeywords}
    University entrance examination, Simulation, Linking, Item Response Theory
\end{IEEEkeywords}

\section{Introduction}
High school students select universities to which they apply considering the difficulty levels of entrance examinations and their academic abilities. Students skip university entrance examinations that are too challenging or too easy given their academic abilities. High schools typically share only the number of students accepted by each university; they do not release the number of unsuccessful students. Thus, we are able to access information about the number of students accepted from different high schools into specific universities, without knowing the names of the students. In this paper, we focus on examination results data in which examinees skip examinations selectively. Our purpose is to compare difficulty levels of such examinations, what we call ``selectively omitted examination data.''

\begin{figure}
    \centering
    \includegraphics[width=\linewidth]{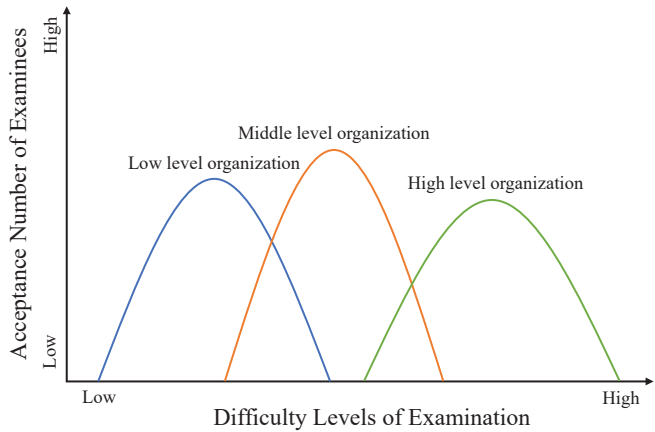}
    \caption{Selectively omitted examination data.}
    \label{fig:fig1}
\end{figure}

Fig. \ref{fig:fig1} presents selectively omitted examination data. It shows the results after being sorted by the difficulty levels of examination. Each organization has a peak – a point at which acceptance is highest. The peak moves to a higher difficulty level of examination when the organization level increases. The organization levels match the difficulty level of examination at the peaks. When the acceptance number of examinees is lower or higher than at the peak, the difficulty levels of examination is lower or higher than the organization level. However, a challenge occurs when the examinations are arranged in random order and the difficulty level is not known as results cannot then be presented by difficulty level.\\
This type of data – the results of certification examinations and the outcome of students' job-hunting activities – is often archived. Universities usually track students' job-hunting activities and note which company the students enter. However, no study has focused on this type of data. If we know the difficulty level of these examinations, we can obtain a new index to assess the organizations' standards, pass rates and difficulty levels of examinations. This index would reflect outcomes of student education from an examination perspective.

Methods such as Classical Test Theory (CTT) \cite{Naovick1966} have been proposed to compare difficulty levels of examinations or their items. Item Response Theory (IRT) is an example of one of the methods from this approach \cite{Hambleton1991}. IRT suggests that the relationship between the accuracy rate of each item and the ability level of a person can be expressed as a function; we can estimate the difficulty levels of different examinations and their items based on the function.

Many studies have connected different tests using those methods; referred to as ``linking'' \cite{Liu2007,Kolen2014,Orlando2000,Sireci1997,Kim1998}. Liu and Walker \cite{Liu2007} demonstrated a connection between the National Assessment of Educational Progress \cite{InstituteofEducationSciences}, the International Assessment of Educational Progress \cite{Mead1995}, the Armed Services Vocational Aptitude Battery \cite{MilitaryAdvantage}, and the North Carolina End-of-Grade Tests \cite{NorthCarolina}. Kolen and Brennan \cite{Kolen2014} connected the American College Testing \cite{ActInc}, and the Iowa Tests of Educational Development \cite{UniversityIowa}.

One of the prerequisites of IRT and CTT is that examinees are challenged to solve all examination items, and IRT and CTT need data regarding who can or cannot solve which items. However, our target data does not meet the prerequisites of IRT and CTT. The examinees skip examinations selectively, and we cannot know which examinations are skipped and why examinees skip them; the examinations are either too easy or too difficult. Thus, we propose a novel method, Peak Shift Estimation, named after the shapes of Fig. \ref{fig:fig1}.

First, we introduced the algorithm of Peak Shift Estimation. Then, we applied Peak Shift Estimation to simulation data to verify its accuracy and robustness. Finally, we applied Peak Shift Estimation to real data from the Tokyo metropolitan area.\\
\section{PEAK SHIFT ESTIMATION}
The algorithm of Peak Shift Estimation is shown below. Considering readability, we described examinations as university entrance examinations and examinees as high schools. We designated the total number of high schools by $m$, the total number of universities by $n$, loop count by $i$, a set of universities as $U_i$, and a set of high schools as $H_i$. As a precondition, we know the top universities of the difficulty level ranking and use them as $U_1$ at Step (0); Tokyo University corresponds to $U_1$ in Japan.

First, Peak Shift Estimation standardizes input data considering the difference in total student numbers between high schools and total acceptance number between universities.

\begin{itemize}
    \item Standardizing step (1) Divide each acceptance number by the corresponding total number of high school students.
    \item Standardizing step (2) Divide each number from step (1) by the total number of universities.
\end{itemize}

Then, Peak Shift Estimation estimates university entrance examination ranks.

\begin{itemize}
    \item Estimating step (0) Rank $U_1$ number one of temporary difficulty level ranking.
\end{itemize}

Loop Estimating step (1) – (6)

\begin{itemize}
    \item Estimating step (1) Sort high schools by summation of acceptance rate of $U_i$ in descending order.
    \item Estimating step (2) Define the top $\lfloor |U_i| \times m \div n \rfloor$
          high schools of the (1) ranking as $H_i$.
    \item Estimating step (3) Cluster universities that are not ranked based on the acceptance rate of $H_i$ with X-means \cite{Pelleg2000}.
    \item Estimating step (4) Sort the university clusters by an average acceptance rate of $H_i$  in descending order.
    \item Estimating step (5) Define the top university cluster of the step (4) ranking as $U_{i+1}$ and rank $U_{i+1}$ as $i+1$ of temporary difficulty level ranking.
    \item Estimating step (6) When all universities are ranked, stop the algorithm.
\end{itemize}

Peak Shift Estimation uses X-means and the temporary difficulty level ranking is stochastically variable. We calculated temporary difficulty level ranking more than once, taking an average of them. Then, we sorted the average and defined them as estimating difficulty level ranking.

\section{Dataset}
Table 1 shows the real data of the Tokyo metropolitan area provided by an investigation firm. It includes the number of students at each high school, the number of high schools, the number of students accepted at each high school and each university entrance examination, number of universities, number of students admitted to each university, and difficulty level of university entrance examinations. We treated the difficulty level of university entrance examinations that students passed as indicative of students' academic ability.

The investigation firm conducted their original examinations, evaluated each student's academic ability, surveyed which university students are accepted, and estimated the difficulty level of university entrance examinations. We treated this as a true ranking.

\begin{figure}[tbp]
    \centering
    \includegraphics[width=\linewidth]{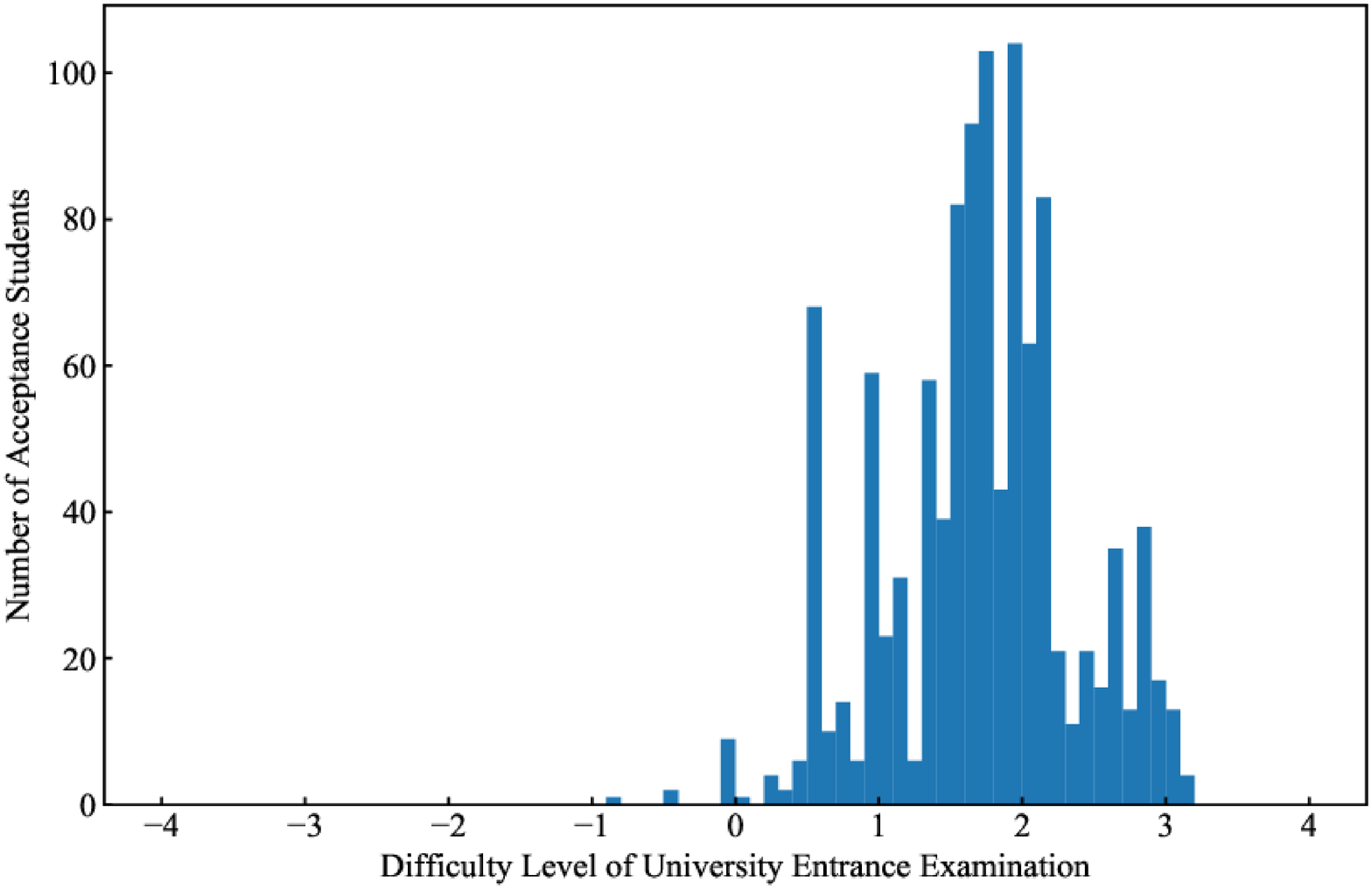}
    \caption{Example of high-level high school.}
    \label{fig:fig2}
\end{figure}

\begin{figure}[tbp]
    \centering
    \includegraphics[width=\linewidth]{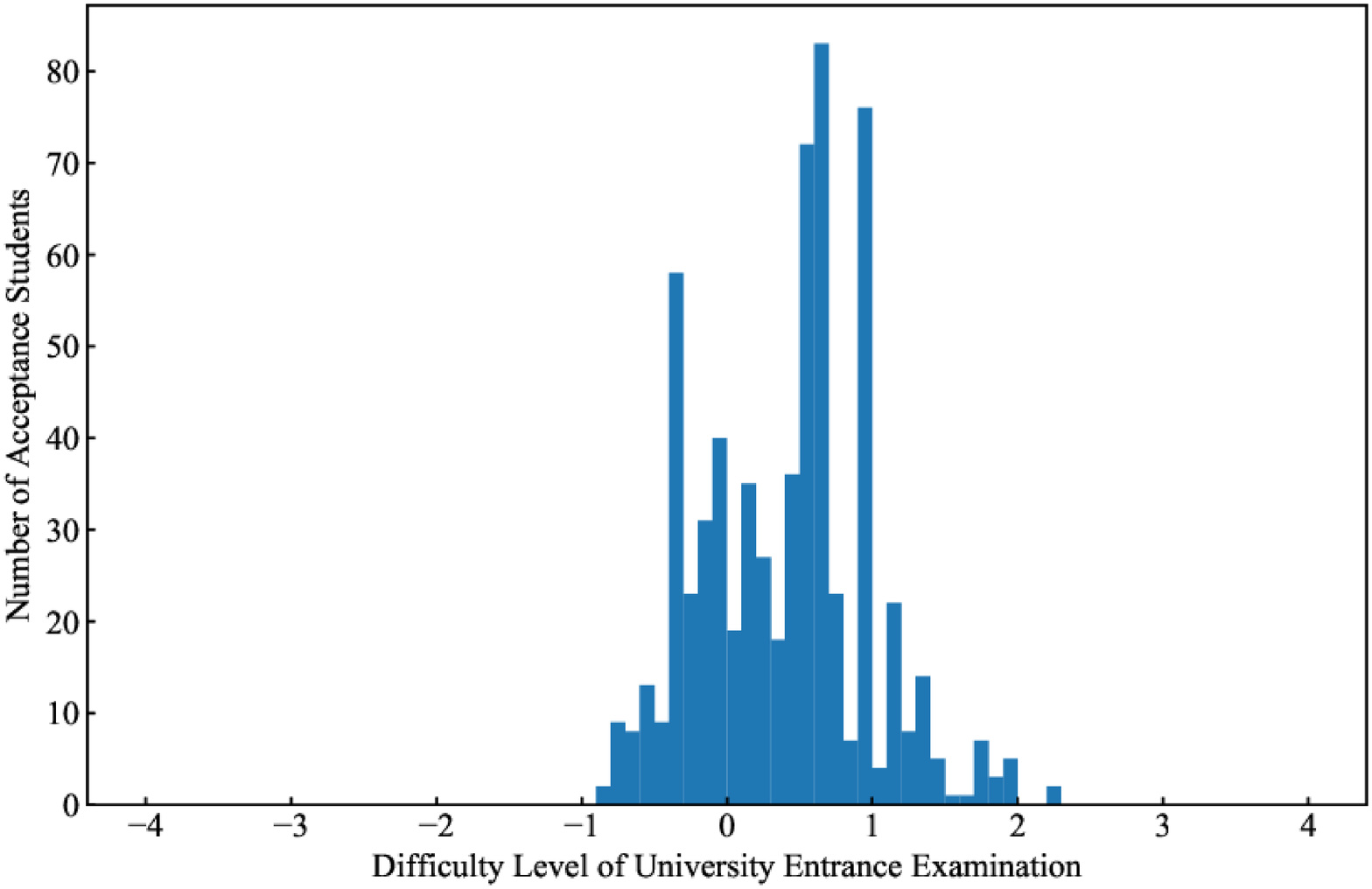}
    \caption{Example of middle-level high school.}
    \label{fig:fig3}
\end{figure}

\begin{figure}[tbp]
    \centering
    \includegraphics[width=\linewidth]{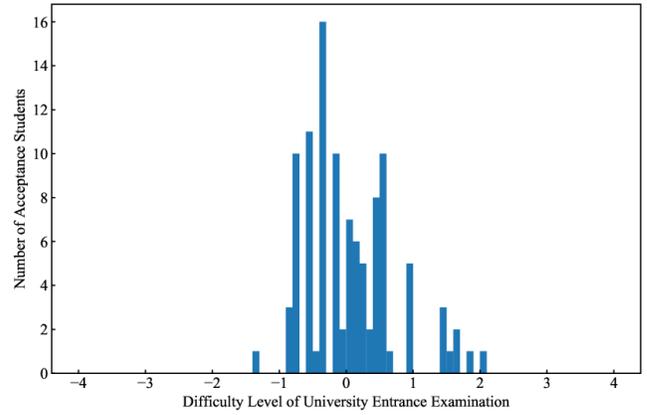}
    \caption{Example of low-level high school.}
    \label{fig:fig4}
\end{figure}

Figs. \ref{fig:fig2}, \ref{fig:fig3}, and \ref{fig:fig4} show examples of the acceptance number of each high school; the peaks move lower as the levels of high schools get lower.

\begin{table*}
    \caption{Summary of datasets}
    \centering
    \begin{tabular}{lrrc}\hline
        Item                                                                    & \multicolumn{1}{l}{Real Data} & \multicolumn{1}{l}{Simulation} & \multicolumn{1}{l}{Formula} \\\hline\hline
        Number of high school students                                          & 283,422                       & 286,000                        & -                           \\
        $\mu$ of students' academic ability distribution                        & -                             & 0                              & -                           \\
        $\sigma$ of students academic ability distribution                      & -                             & 1                              & -                           \\
        Number of high schools                                                  & 1,078                         & 1,100                          & k                           \\
        Number of students at each high school                                  & Ave. 262.9                    & 260                            & m                           \\
        $\mu$ of high schools average academic ability distribution             & 0.3                           & 0                              & -                           \\
        $\sigma$ of high schools average academic ability distribution          & 0.64                          & 0.8                            & $\sigma_a$                  \\
        $\sigma$ of students academic ability in each high school               & 0.61                          & 0.6                            & $\sigma_e$                  \\
        $\sigma$ of $\sigma$ of students academic ability in each high school   & 0.20                          & -                              & -                           \\
        Limitation of $\sigma$ of students academic ability in each high school & -                             & 1.96 $\times$ 0.20             & -                           \\
        Number of universities                                                  & 157                           & 160                            & -                           \\
        Number of entrance students in each university                          & Ave. 1555.7                   & 1,600                          & -                           \\
        $\mu$ of the difficulty levels of entrance examinations distribution    & -                             & 0                              & -                           \\
        $\sigma$ of the difficulty levels of entrance examinations distribution & -                             & 1                              & -                           \\
        Limitation of entrance examination ability                              & -                             & 1                              & -                           \\\hline
    \end{tabular}
    \label{tab:table1}
\end{table*}


\section{SIMULATION DATA}

\subsection{Generating Data}
We applied Peak Shift Estimation to simulation data to verify the accuracy and robustness of the model. Data were generated based on table \ref{tab:table1}.

\subsubsection{The Procedure of Generating Data}\ \\
The procedure of high school data generation was as follows:

\begin{itemize}
    \item High school data step (1)	Generate number of high school students and allocate students' academic ability following a normal distribution.
    \item High school data step (2)	Generate number of high schools and allocate high schools' average academic ability to them following a normal distribution.
    \item High school data step (3)	Assign high school students stochastically to high schools whose average academic ability is close to students' academic ability, considering the number of students in each high school.
    \item High school data step (4)	Calculate the standard deviation of students' academic abilities for each high school. When at least one high schools' standard deviation is over the limitation, go back to Step (3).
    \item High school data step (5)	Rank the high schools based on the average of their students' academic abilities.
\end{itemize}

The procedure for university data generation is given below:

\begin{itemize}
    \item University data step (1) Generate universities and allocate difficulty levels of examinations to them following a normal distribution.
    \item University data step (2) Each high school student chooses universities to take entrance examinations considering that the difficulty levels of examinations are not over or under the student's academic ability $\pm$ limitation of entrance examination ability. Each high school student chooses any number of universities and becomes a candidate for chosen universities.
    \item University data step (3) In the order corresponding to difficulty levels of examination, each university accepts candidate students up to the number of university students in the order corresponding to the academic ability of candidate students. Universities cannot accept students who have been accepted by the other universities already. Students become entrance students at the universities that accepted them.
    \item University data step (4) Each university judges the candidate students who have higher academic ability as acceptance students, rather than the entrance students with the lowest academic ability among them.
\end{itemize}

\subsubsection{Parameters of Generating Data}\ \\
Table \ref{tab:table1} shows simulation parameters. Based on students' academic ability in each high school, we can calculate $\sigma$ of students' academic ability in each high school and $\sigma$ of high schools' average academic ability. When we used these values as simulation parameters, the simulation could not meet the requirement of the high school data step (4). We were concerned about the lack of student data; thus, we set up the relational expression between $\sigma$ of students' academic ability in each high school and $\sigma$ of high schools' average academic ability.

Define each high school as $i$, each student in each high school as $j$, distribution of student academic ability as $ \mathcal{N}(0,\,1^{2})\, $, distribution of high schools' average academic ability as $ \mathcal{N}(0,\,\sigma^{2}_a)\, $, distribution of students' academic ability in each high school as $ \mathcal{N}(\sigma_i\sigma_a,\,\sigma^{2}_e)\, $, $a_i$ as constants of each high school, $b_{i,j}$ as constants of each student, and each student's academic ability as $x_{i,j}=a_i\sigma_a+b_{i,j}\sigma_e$.

From $ \mathcal{N}(0,\,1^{2})\, $: distribution of high schools' average academic ability,

\begin{equation}
    0=\frac{1}{km}\sum_i^k\sum_j^m\left(a_{i}\sigma_{a}+b_{i,j}\sigma_{e}\right).
\end{equation}
\begin{equation}
    1^{2}=\frac{1}{km}\sum_i^k\sum_j^m\left(a_{i}\sigma_{a}+b_{i,j}\sigma_{e}\right)^{2}.
\end{equation}

From $ \mathcal{N}(0,\,\sigma^{2}_a)\, $: distribution of high schools' average academic ability,

\begin{equation}
    0=\frac{1}{k}\sum_i^k\left(a_{i}\sigma_{a}\right).
\end{equation}
\begin{equation}
    \sigma_{a}^{2}=\frac{1}{k}\sum_i^k\left(a_{i}\sigma_{a}\right)^{2}.
\end{equation}

From (3),

\begin{equation}
    \sum_i^ka_{i}=0.
\end{equation}

From (4),

\begin{equation}
    \sum_i^ka_i^2=k.
\end{equation}

From $ \mathcal{N}(a_i\sigma_{a},\,\sigma^{2}_e)\, $: distribution of students' academic ability in each high school,

\begin{equation}
    a_{i}\sigma_{a}=\frac{1}{m}\sum_j^m\left(a_i\sigma_{a}+b_{i,j}\sigma_{e}\right).
\end{equation}
\begin{equation}
    \sigma_{e}^2=\frac{1}{m}\sum_j^m\left(b_{i,j}\sigma_{e}\right)^2.
\end{equation}

From (7),

\begin{equation}
    \sum_j^mb_{i,j}=0.
\end{equation}

From (8),

\begin{equation}
    \sum_j^mb_{i,j}^2=m.
\end{equation}

From (2),(5),(6),(9),(10),

\begin{equation}
    1^2=\sigma_a^2+\sigma_e^2.
\end{equation}

Then, we put the sigmae and sigmaa of simulation as below.

\begin{equation}
    \sigma_e=0.6.
\end{equation}
\begin{equation}
    \sigma_a=0.8.
\end{equation}

\subsection{Generating Data}
\begin{figure}[tbp]
    \centering
    \includegraphics[width=\linewidth]{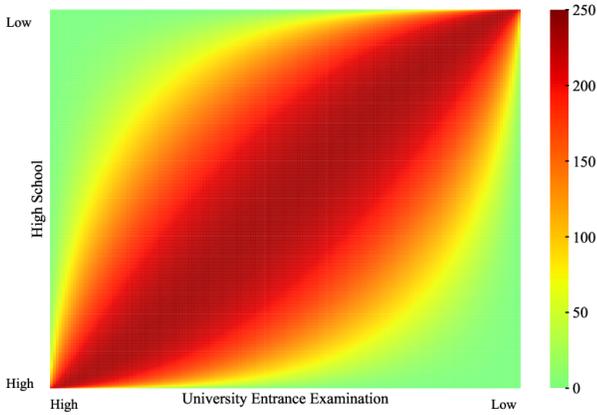}
    \caption{Number of candidates.}
    \label{fig:fig5}
\end{figure}

We generated data by simulation 1,000 times. Fig. \ref{fig:fig5} shows the candidate students' heatmap between universities and high schools. The candidate students are distributed between two arcs.

\begin{figure}[tbp]
    \centering
    \includegraphics[width=\linewidth]{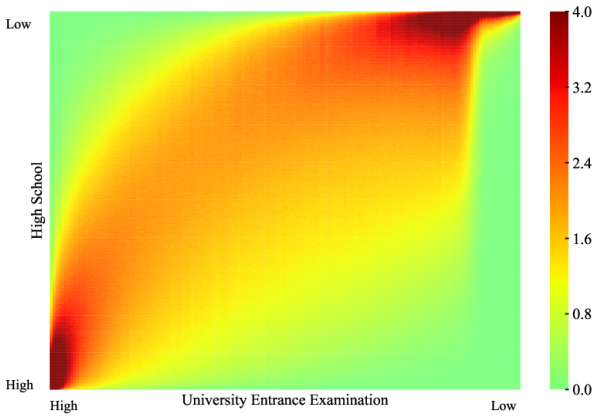}
    \caption{Number of entrance places.}
    \label{fig:fig6}
\end{figure}

Fig. \ref{fig:fig6} shows the entrance students' heatmap between universities and high schools. The students of top and bottom ranked universities are concentrated in the top and bottom ranked high schools. The students at the middle-rank universities are distributed between two arcs; they enter the highest university of those whose examinations they passed.

The standard deviations of high school students' academic abilities and the difficulty levels of examinations are the same, however, the number of students is far greater than the number of universities. Thus, the high school students' academic abilities are more widely distributed than the difficulty levels of examinations Accordingly, some of the highest ranked high school students have no choice but to enter the top rank universities, even, those whose difficulty levels are lower than their academic abilities; the students at the top ranked universities are concentrated in the top ranked high schools. High school students enter a university in the order corresponding to the difficulty levels; the upper arc is denser than the lower arc. As the lowest ranked universities lose their candidate students, the lower arc suddenly rises.

\begin{figure}[tbp]
    \centering
    \includegraphics[width=\linewidth]{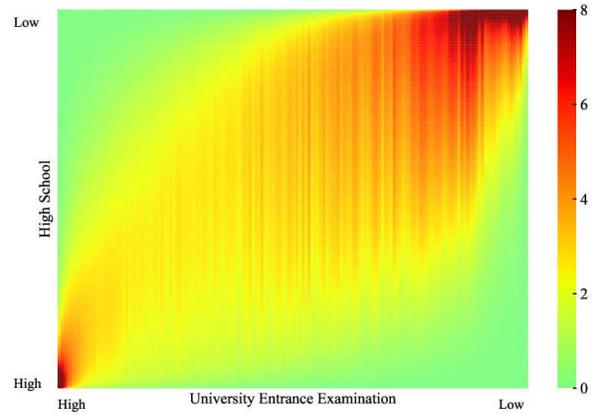}
    \caption{Acceptance numbers.}
    \label{fig:fig7}
\end{figure}

Fig. \ref{fig:fig7} shows the accepted students' heatmap between universities and high schools. The students at top and bottom rank universities are concentrated in the top and bottom ranked high schools. The students at the middle-rank universities are distributed between two arcs. Each university judges the candidate students who have higher academic ability as successful students, in comparison to the lowest academic ability of entrance students. Accordingly, the students who are between the two arcs of Fig. \ref{fig:fig5} and higher than the lower arc of Fig. \ref{fig:fig6}, become successful students.


\subsection{Applying Peak Shift Estimation}

We applied Peak Shift Estimation to the simulation data, calculated temporary difficulty level ranking 1,000 times, and took an average of the values; we sorted the average and defined them as the difficulty level ranking. The simulation data includes 1,000 pattern results and we obtained 1,000 pattern difficulty level rankings.\\
Spearman's rank correlation coefficient was calculated between the difficulty level ranking, ``estimated ranking'', and the ranking of the difficulty levels of examinations – ``true ranking''. To verify the robustness of Peak Shift Estimation, we dropped some high schools, randomly changing the rate of dropped high schools as a proportion of all high schools from 0.0 to 0.9. This scenario assumes that some organizations' data are missing.

\begin{figure}
    \centering
    \includegraphics[width=\linewidth]{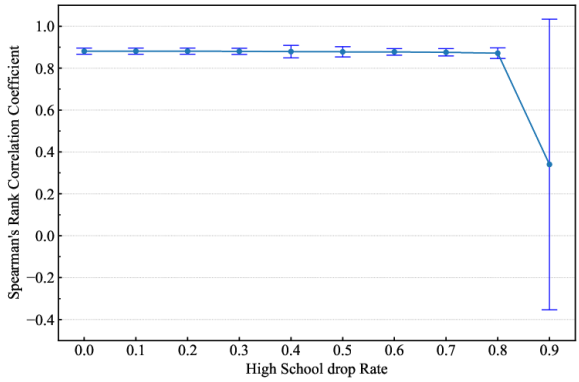}
    \caption{Rank correlation coefficient of simulation results.}
    \label{fig:fig8}
\end{figure}

Fig. \ref{fig:fig8} shows the result of the simulation. The rank correlation coefficients are around 0.88 and stable when the dropped high school ratio changes from 0.0 to 0.8. This result suggests that Peak Shift Estimation estimates the rank order of difficulty level of university entrance examinations.

\begin{figure}
    \centering
    \includegraphics[width=\linewidth]{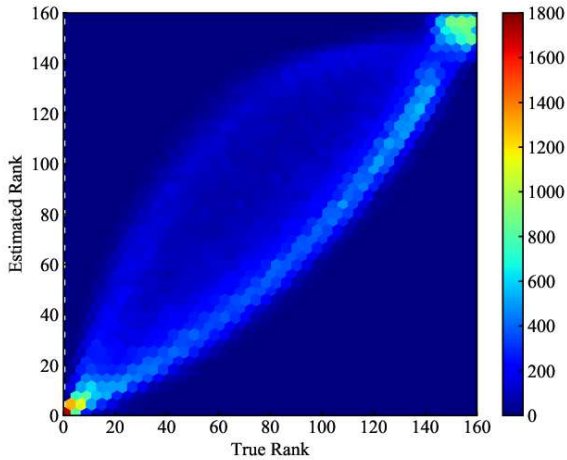}
    \caption{Heatmap between true and estimated ranks.}
    \label{fig:fig9}
\end{figure}

Fig. \ref{fig:fig9} shows the heatmap between the true ranks and estimated ranks. The top and bottom ranked universities are estimated with high accuracy. The middle-rank universities are distributed in two arcs, an upper and a lower arc; the lower arc is more dense and gradual than the upper arc. The ranks of the universities on the upper arc are estimated as lower than their true ranks; the ranks of the universities on the lower arc are estimated as higher than their true ranks. Peak Shift Estimation estimates universities' ranks in descending order. This means that some universities' ranks are estimated lower and other universities' ranks are pushed out and become higher.

\begin{figure}
    \centering
    \includegraphics[width=\linewidth]{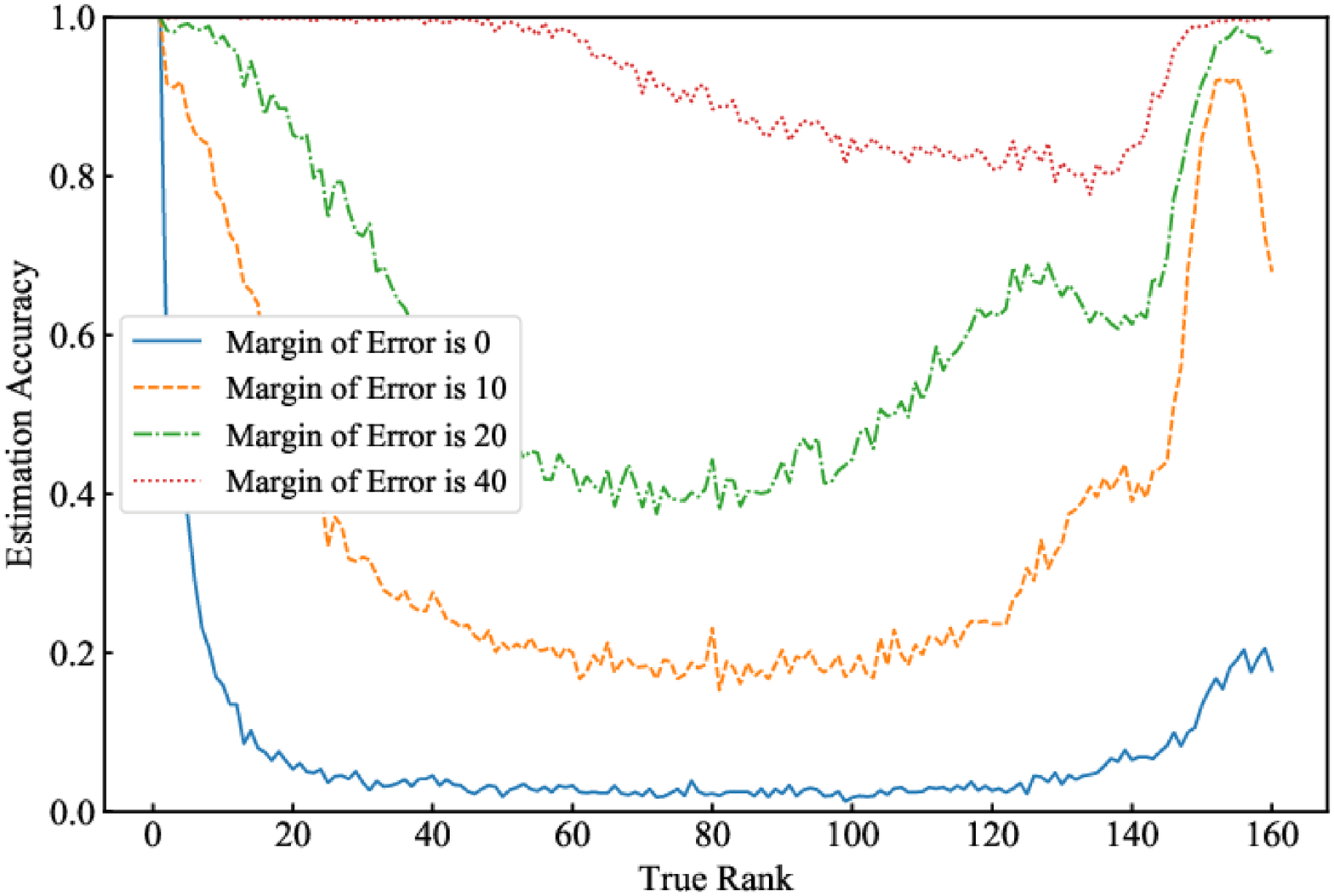}
    \caption{Accuracy of estimated ranks at each true rank.}
    \label{fig:fig10}
\end{figure}

Fig. \ref{fig:fig10} shows the accuracy of estimated ranks at each true rank. We changed the margin of error: 0, 10, 20, 40 ranks. All lines start at high accuracy, drop on the middle ranks, and rise up on the bottom ranks. Only the line for the margin of error of 40 maintains 0.8 accuracy or above. The difficulty levels of examinations are generated following a normal distribution. The number of middle-rank universities is large, and their difficulty levels are very close. For this reason, Peak Shift Estimation cannot estimate ranks of the middle-rank universities precisely.


\subsection{Discussion}
We showed that Peak Shift Estimation very robustly estimates rank order of difficulty level of university entrance examinations. However, the accuracy of estimated ranks is not so high, especially for the middle ranks. This is because the number of middle-rank universities is large, and the difficulty levels are very close. Some universities' ranks are estimated lower and other universities' ranks are pushed out and become higher.

Only the line of margin of error of 40 maintains 0.8 accuracy or higher in Fig. \ref{fig:fig10}. From this result, Peak Shift Estimation would be suitable for estimating the multi-level scale of universities: A, B, C, and D rank university entrance examination.

\section{REAL DATA}

\subsection{Applying Peak Shift Estimation}

We applied Peak Shift Estimation to the acceptance number of each high school for each university in the Tokyo metropolitan area and compared the result with the difficulty level ranking created by the investigation firm. The investigation firm has conducted nationwide practice entrance examinations and tracked which university entrance examinations the examinees passed. Based on this survey, the investigation firm ranks the university entrance examinations. We treated this as a true ranking.

We calculated the temporary difficulty level ranking 1,000 times and took an average of the results. Then, we sorted the average of the results and defined them as the difficulty level ranking, ``estimated ranking.'' The Spearman's rank correlations coefficient between the estimated ranking and the true ranking was 0.844.

\begin{figure}
    \centering
    \includegraphics[width=\linewidth]{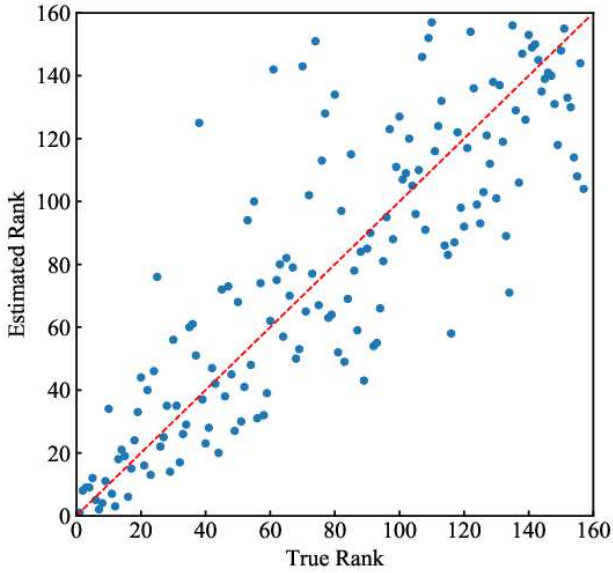}
    \caption{Results of real data.}
    \label{fig:fig11}
\end{figure}

Fig. \ref{fig:fig11} shows the true ranks and estimated ranks; each point shows each university entrance examination and the red dashed line shows where a true rank equals an estimated rank. The results are scattered between two arcs like Fig. \ref{fig:fig9}; universities on the top and bottom ranks are estimated more accurately than universities in the middle ranks. One university at 38 true rank was estimated at 133 rank; this university is the top art college and its entrance examination differs vastly from other universities. Thus, Peak Shift Estimation was significantly inaccurate in estimating the rank of the university.

\begin{figure}
    \centering
    \includegraphics[width=\linewidth]{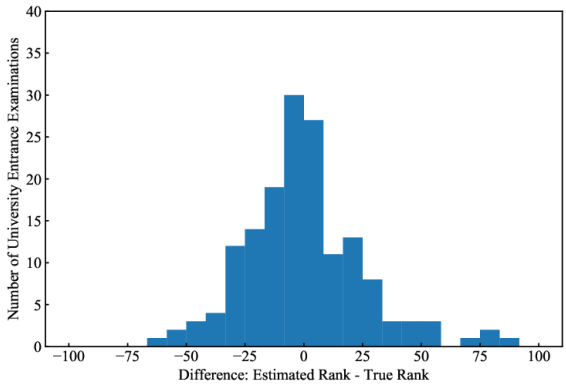}
    \caption{Difference between estimated rank and true rank.}
    \label{fig:fig12}
\end{figure}

Fig. \ref{fig:fig12} shows a histogram of the difference between the true ranks and the estimated ranks. The histogram's shape is nearly symmetric; the estimated ranks are equally missed on both upper and lower sides. About 80 percent of universities' difference is within 25 ranks.


\subsection{Cause of Decreasing Estimation Accuracy}
To investigate the cause of decreasing estimation accuracy, we validated how one university entrance examination influences the whole estimated ranking. In addition, we analyzed how Peak Shift Estimation clustered the university entrance examinations.

\subsubsection{Dropping universities one by one}
\begin{figure}
    \centering
    \includegraphics[width=\linewidth]{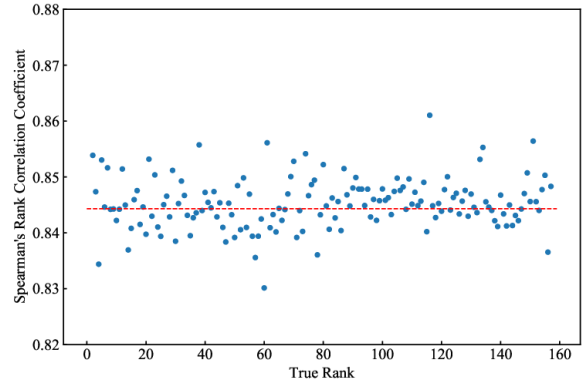}
    \caption{Rank correlation coefficient of dropping each university at each estimated rank.}
    \label{fig:fig13}
\end{figure}

\begin{figure}
    \centering
    \includegraphics[width=\linewidth]{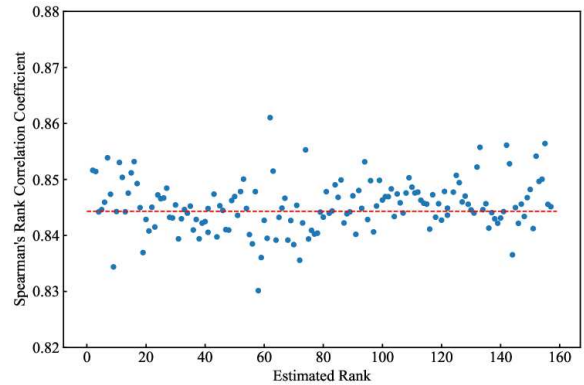}
    \caption{Rank correlation coefficient of dropping each university at each true rank.}
    \label{fig:fig14}
\end{figure}

\begin{figure}
    \centering
    \includegraphics[width=\linewidth]{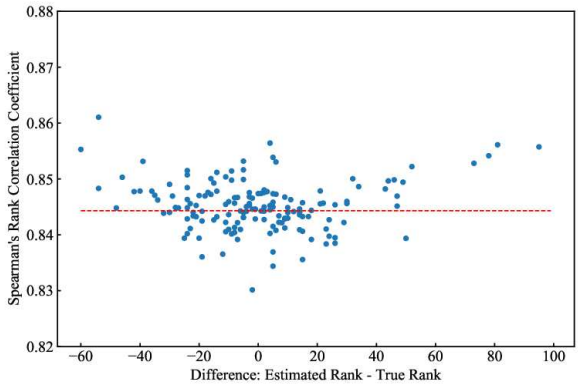}
    \caption{Rank correlation coefficient of dropping each university at each difference.}
    \label{fig:fig15}
\end{figure}

\ \\
We dropped a university one by one from the real data and applied Peak Shift Estimation to it. Figs. \ref{fig:fig13}, \ref{fig:fig14}, and \ref{fig:fig15} show the results; horizontal axes show the estimated rank, the true rank, and the difference between the true rank and the estimated rank of the dropped university. The red dashed lines show where rank correlation coefficient equals 0.844, the rank correlation coefficient before dropping. Table \ref{tab:table2} shows the summary of the top five universities in Spearman's rank correlation coefficient.

Peak Shift Estimation determines a difficulty level ranking from the top rank with X-means; Fig. \ref{fig:fig13} and table \ref{tab:table2} show, however, that even universities at low estimated ranks decrease the estimation accuracy, e.g., when the university whose estimated rank is 155 was dropped, the rank correlation coefficient went up to 0.856.

We could not find any relationship between the estimated ranks, the true ranks, and the difference, and the rank correlation coefficients.\\


\begin{table}
    \caption{Summary of the top five universities in rank correlation coefficient.}
    \centering
    \begin{tabular}{lrrrr}
        \hline
        \multicolumn{1}{c}{University}                   &
        \multicolumn{1}{l}{\begin{tabular}{c} Rank \\ correlation \\ coefficient \end{tabular}} &
        \multicolumn{1}{l}{\begin{tabular}{c} Estimated \\ rank \end{tabular}} &
        \multicolumn{1}{l}{\begin{tabular}{c} True \\ rank \end{tabular} }&
        \multicolumn{1}{l}{Diff}
        \\ \hline\hline
        \begin{tabular}{c} dropped \\ university 1\end{tabular}& 0.861 & 62  & 116 & -54 \\
        \begin{tabular}{c} dropped \\ university 2\end{tabular} & 0.856 & 155 & 151 & 4   \\
        \begin{tabular}{c} dropped \\ university 3\end{tabular}& 0.856 & 142 & 61  & 81  \\
        \begin{tabular}{c} dropped \\ university 4\end{tabular}& 0.856 & 133 & 38  & 95  \\
        \begin{tabular}{c} dropped \\ university 5\end{tabular}& 0.855 & 74  & 134 & -60 \\ \hline
    \end{tabular}
    \label{tab:table2}
\end{table}

\subsubsection{Clustering network}
\begin{figure}
    \centering
    \includegraphics[width=\linewidth]{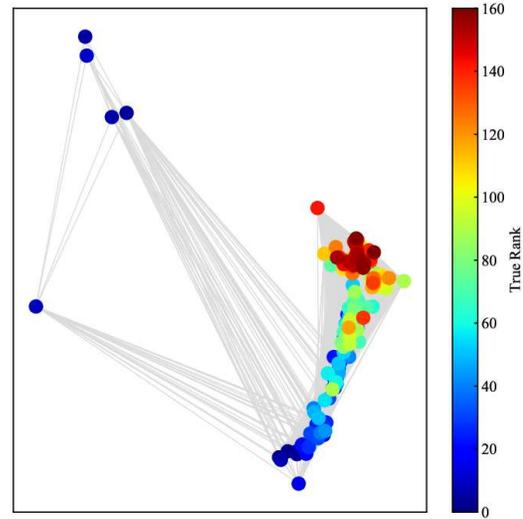}
    \caption{Graph of clustering.}
    \label{fig:fig16}
\end{figure}

\ \\
Fig. \ref{fig:fig16} shows the network of clustering by X-means; each node means each university. When universities are clustered into the same group, they are connected by edges. Each edge has the number being clustered into the same group, from 1 to 1000. Universities are located from lower to upper in descending order. They build small communities, e.g., upper-left four universities.

We cut edges gradually from the edges that have a lower number being clustered into the same group, and searched communities by Girvan-Newman community detection algorithm \cite{Girvan2002}. We found several communities and some of them have the same features unrelated to the difficulty of entrance examinations.

\begin{figure}
    \centering
    \includegraphics[width=\linewidth]{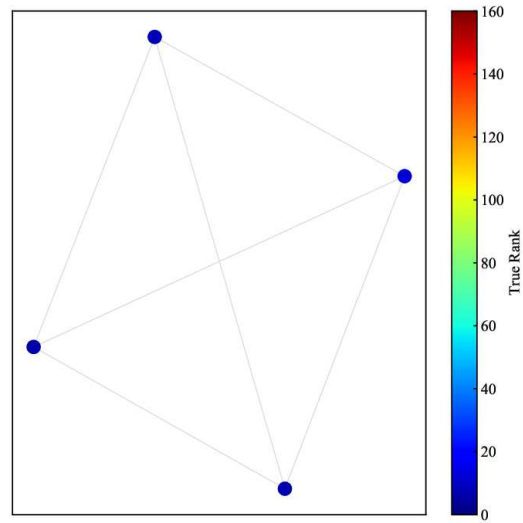}
    \caption{Medical colleges cluster.}
    \label{fig:fig17}
\end{figure}

Fig. \ref{fig:fig17} shows the medical colleges cluster; all four universities are medical colleges. In Japan, medical departments are perceived as special because they require high academic ability and very expensive academic fees as they are stepping-stones to medical licenses. Some high schools have special courses to facilitate entrance to medical colleges. Only elite high school students take entrance examinations for medical colleges and they tend not to take other departments' entrance examinations. This tendency supports the medical college cluster.

\begin{figure}
    \centering
    \includegraphics[width=\linewidth]{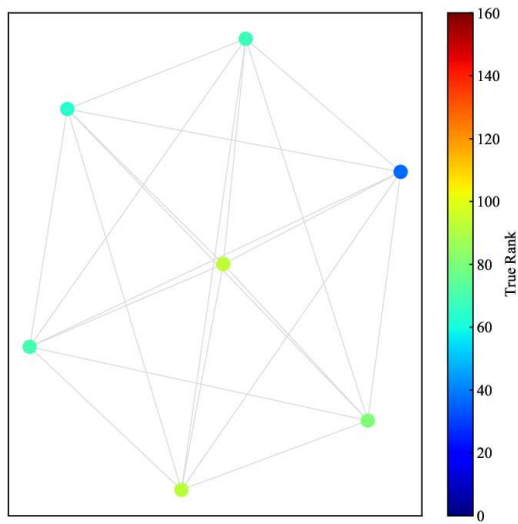}
    \caption{Women's universities cluster.}
    \label{fig:fig18}
\end{figure}

Fig. \ref{fig:fig18} shows the women's universities cluster; all seven universities are women's universities. Table \ref{tab:table3} shows the summary of the women's universities cluster. A wide variety of universities, from 37 true rank to 93 true rank comprise this cluster. However, their estimated ranks are very close, from 50 to 57. This means that they are estimated as the same level, even though their true ranks are different.

Girls' schools are popular in Japan, and some girls select girls' schools from kindergarten to university. Table \ref{tab:table4} shows the acceptance number of coeducational and girls' schools. The coeducational high schools include both male and female students; we estimated half of the students are female and recalculated the rate (table \ref{tab:table4}). The women's college acceptance rate per girls' high school students is much higher than the rate of women's colleges acceptance for coeducational high school students. This tendency would build the women's universities cluster.


\begin{table}
    \caption{Summary of women's universities cluster}
    \centering
    \begin{tabular}{lrr}\hline
        \multicolumn{1}{l}{University}     &
        \multicolumn{1}{l}{Estimated rank} &
        \multicolumn{1}{l}{True rank}                \\
        \hline\hline
        women's university 1               & 50 & 68 \\
        women's university 2               & 51 & 37 \\
        women's university 3               & 52 & 92 \\
        women's university 4               & 53 & 69 \\
        women's university 5               & 54 & 81 \\
        women's university 6               & 55 & 64 \\
        women's university 7               & 57 & 93 \\
        \hline
    \end{tabular}
    \label{tab:table3}
\end{table}

\begin{table}
    \caption{Acceptance number of coeducational and girls' schools}
    \centering
    \begin{tabular}{llrr}\hline
        \multicolumn{1}{l}{High\par School} &
        \multicolumn{1}{l}{University}  &
        \multicolumn{1}{l}{Acceptance\par Number}                              &
        \multicolumn{1}{l}{Rate}                                           \\
        \hline\hline
        Coeducational                   & Coeducational & 252,023  & 0.930 \\
        Coeducational                   & Girl's        & 190,89   & 0.070 \\
        \hline
        Girl's                          & Coeducational & 32,258   & 0.789 \\
        Girl's                          & Girl's        & 8,631    & 0.211 \\
        \hline
        Coeducational*                   & Coeducational* & 126011.5* & 0.868* \\
        Coeducational*                   & Girl's*       & 190,89*   & 0.131* \\
        \hline
    \end{tabular}
    \label{tab:table4}
    \\$\ast$ After recalculation
\end{table}

\subsection{Discussion}
We applied Peak Shift Estimation to the real data of the Tokyo metropolitan area. The rank correlations coefficient between the estimated ranking and the true ranking was 0.844, and 80 percent of universities' difference is within 25 ranks. Then, we found that one university would decrease the entire accuracy of the estimation regardless of its true rank. We showed universities build several communities that have the same external factors, e.g., medical departments and girls' schools, that would decrease the accuracy of estimation.

\section{CONCLUSION}
We proposed Peak Shift Estimation to estimate the rank of examination difficulty levels when examinees skip examinations selectively, and verified the accuracy and robustness of Peak Shift Estimation using university entrance examinations as an example.

First, we applied Peak Shift Estimation to simulation data and showed that Peak Shift Estimation very robustly estimates the rank order of difficulty level of university entrance examinations. However, we found the accuracy of estimated ranks was lower, especially for the middle ranks. This is because the number of middle-rank universities is large, and the difficulty levels are very close. Some universities' ranks were estimated higher and others' ranks were pushed out and became lower. Only a margin of error of 40 maintained a 0.8 accuracy level or higher. \\
From this result, Peak Shift Estimation would be suitable for estimating the multi-level scale of universities: A, B, C, and D rank university entrance examinations.

We applied Peak Shift Estimation to the real data of the Tokyo metropolitan area and showed the rank correlations coefficient between the estimated ranking and the true ranking was 0.844 and 80 percent of universities' difference is within 25 ranks.\\
Then, we found that one university would decrease the accuracy of the entire estimation regardless of its true rank, and universities build several communities that share the same external factors, e.g., medical departments and girls' schools, which also decrease the accuracy of estimation.\\
The estimation accuracy of Peak Shift Estimation is low and requires improvement. However, this is the first study to focus on selectively omitted examination data, and this research contributes to understanding the data.

\bibliographystyle{unsrt}
\bibliography{references}

\begin{thebibliography}{10}

\bibitem{Naovick1966}
Melvin~R. Novick.
\newblock The axioms and principal results of classical test theory.
\newblock {\em Journal of mathematical psychology}, 3(1):1--18, 1966.

\bibitem{Hambleton1991}
Ronald~K. Hambleton, H.~Swaminathan, and H.~Jane Rogers.
\newblock {\em Fundamentals of Item Response Theory}, volume~2.
\newblock {Sage Press}, Newbury Park, CA, 1991.

\bibitem{Liu2007}
Jinghua Liu and Michael~E. Walker.
\newblock Score linking issues related to test content changes.
\newblock In {\em Linking and Aligning Scores and Scales}, pages 109--134.
  Springer, 2007.

\bibitem{Kolen2014}
Michael~J. Kolen and Robert~L. Brennan.
\newblock {\em Test Equating, Scaling, and Linking: Methods and Practices}.
\newblock Springer, New York, NY, 2014.

\bibitem{Orlando2000}
Maria Orlando, Cathy~D. Sherbourne, and David Thissen.
\newblock Summed-score linking using item response theory: Application to
  depression measurement.
\newblock {\em Psychological Assessment}, 12(3):354--359, 2000.

\bibitem{Sireci1997}
Stephen~G Sireci.
\newblock Problems and issues in linking assessments across languages.
\newblock {\em Educational Measurement: Issues and Practice}, 16(1):12--19,
  1997.

\bibitem{Kim1998}
Seock-Ho Kim and Allan~S. Cohen.
\newblock A comparison of linking and concurrent calibration under item
  response theory.
\newblock {\em Applied Psychological Measurement}, 22(2):131--143, 1998.

\bibitem{InstituteofEducationSciences}
{Institute of Education Sciences}.
\newblock {"The Nation's Report Card: NAEP."} accessed feb. 13, 2021.
\newblock \url{https://nces.ed.gov/nationsreportcard/}.

\bibitem{Mead1995}
National~Research Council.
\newblock International assessment of educational progress.
\newblock In {\em International Comparative Studies in Education: Descriptions
  of Selected Large-Scale Assessments and Case Studies}, pages 48--57. The
  National Academies Press, Washington, DC, 1995.

\bibitem{MilitaryAdvantage}
{Military Advantage}.
\newblock {"The ASVAB Test."} accessed feb. 13, 2021.
\newblock \url{https://www.military.com/join-armed-forces/asvab}.

\bibitem{NorthCarolina}
{North Carolina Department of Public Instruction}.
\newblock {"End-of-Grade (EOG)."} accessed feb. 13, 2021.
\newblock
  \url{https://www.dpi.nc.gov/districts-schools/testing-and-school-accountability/}\\\url{state-tests/end-grade-eog}.

\bibitem{ActInc}
{Act, Inc.}
\newblock {"The ACT - Solutions for College and Career Readiness."} accessed
  feb. 13, 2021.
\newblock \url{http://www.act.org}.

\bibitem{UniversityIowa}
{The University of Iowa}.
\newblock {"Iowa Testing Programs."} accessed feb. 13, 2021.
\newblock \url{https://education.uiowa.edu/centers/iowa-testing-programs}.

\bibitem{Pelleg2000}
Dan Pelleg and Andrew~W. Moore.
\newblock X-means: Extending k-means with efficient estimation of the number of
  clusters.
\newblock In {\em ICML '00: Proceedings of the Seventeenth International
  Conference on Machine Learning}, volume~1, pages 727--734. Morgan Kaufmann
  Publishers Inc., 2000.

\bibitem{Girvan2002}
Michelle Girvan and Mark E.~J. Newman.
\newblock Community structure in social and biological networks.
\newblock volume~99, pages 7821--7826. National Acad Sciences, 2002.

\end{thebibliography}
\end{document}